\documentclass{optica-article}

\journal{opticajournal} 

\articletype{Research Article}

\usepackage{lineno}
\usepackage{siunitx}

\usepackage{textcomp}
\usepackage{xcolor}
\usepackage{hyperref}
\hypersetup{
    colorlinks=true,
    linkcolor=blue,
    filecolor=magenta,      
    urlcolor=cyan,
    pdftitle={Overleaf Example},
    pdfpagemode=FullScreen,
    }
\usepackage{orcidlink}
\DeclareSIUnit{\millibar}{mbar}
\begin{document}

\title{Group Delay Dispersion Measurements of Novel Multilayer Interference Coatings in the Mid-Infrared Spectral Regime}

\author{Ulrich Galander,\authormark{1,2,*} Maximilian Prinz,\authormark{1,2} Lukas W. Perner,\authormark{3} and Oliver H. Heckl\authormark{1}}

\address{\authormark{1}Optical Metrology Group, Faculty of Physics, University of Vienna, Boltzmanngasse 5, 1090 Vienna, Austria\\
\authormark{2}Vienna Doctoral School in Physics, University of Vienna, Boltzmanngasse 5, 1090 Vienna, Austria\\
\authormark{3}Laboratoire Temps-Fréquence, Institut de Physique, Université de Neuchâtel, Avenue de Bellevaux 51, 2000 Neuchâtel, Switzerland
}

\email{\authormark{*}ulrich.galander@univie.ac.at} 


\begin{abstract} 
We present the methods and results for broadband group delay dispersion measurements for all-monocrystalline and amorphous-crystalline hybrid supermirrors, as well as an all-amorphous mirror in the wavelength range from \SIrange{2.5}{4.8}{\micro \metre}. Measurements are performed using a custom-built white light interferometer that allows for balanced and unbalanced measurement configurations. We compare the results to theoretical transfer-matrix method simulations and an alternative measurement using a commercial Fourier-transform infrared spectrometer. Additionally, we investigate group delay dispersion by direct differentiation using a local polynomial smoothing approach (Savitzky-Golay filter) and find strong consistency between our results, the theoretical prediction and the estimation using this method.
\end{abstract}


\section{Introduction}
\label{sec:intro}
Highly-reflective mirrors are an indispensable tool for many applications in the mid-infrared spectral region, ranging from precision molecular spectroscopy~\cite{picque_frequency_2019} to gravitational wave detection~\cite{cole_substrate-transferred_2023}.
To validate the properties of such mirrors, it is crucial to characterize their performance, such as reflectivity $R$, transmission $T$, absorption $A$, and scatter $S$ as precisely as possible.
For some applications, especially when using mode-locked laser sources, such as attosecond physics~\cite{hentschel_attosecond_2001}, tight control of dispersive properties is necessary.
In this work, we focus on group delay dispersion (GDD) measurements, using a white light interferometer (WLI) operating in unbalanced configuration. We compare this to measurements in balanced configuration with a commercial Fourier-transform infrared spectrometer (FTIR) to validate the measurement approach. 

\begin{figure}[ht!]
\centering\includegraphics{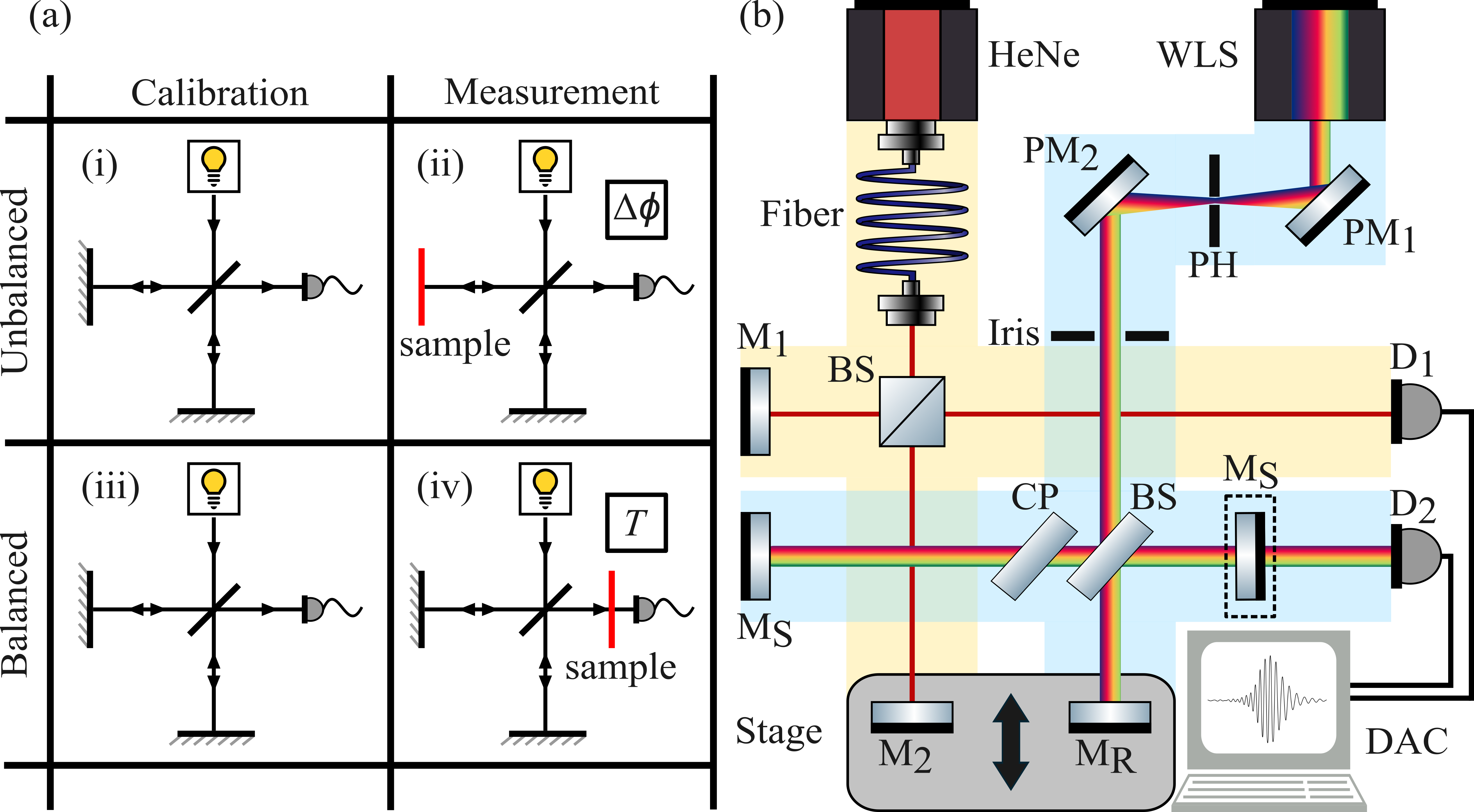}
\caption{(a) The different operation configurations of a Michelson-type interferometer (or FTIR). The red line indicates the sample. (i) \& (iii) Calibration of the setup. It is the same for both operation configurations. (ii) In unbalanced configuration, the sample is placed as one end mirror of the interferometer, and the relative phase difference $\Delta\phi$ between the two arms is measured. (iv) In balanced configuration, the sample is located between the BS and D\textsubscript{2}, and the transmitted intensity $T$ of the sample is measured. (b) The experimental setup. It is divided into two interferometers: one for wavelength calibration (yellow background), and a white light interferometer (blue background). WLS: white light source; PM\textsubscript{1,2}: off-axis parabolic mirrors; PH: precision pinhole; BS: beamsplitter; CP: compensator plate; HeNe: helium-neon laser; M\textsubscript{S}: sample; M\textsubscript{R}: end mirror WLI; M\textsubscript{1,2}: end mirrors reference interferometer; D\textsubscript{1,2}: detector; DAC: data acquisition card.}
\label{fig:setup}
\end{figure}

Balanced and unbalanced configuration refer to different measurement arrangements in a WLI, typically a Michelson-type interferometer. An unbalanced WLI is characterized by only one interferometer arm interacting with a sample, hence, detecting the relative phase shift $\Delta \phi$ between them (see Fig.~\ref{fig:setup}(a)(ii)), whereas in a balanced WLI, no relative phase shift between the arms occurs, but the transmitted intensity $T$ through the sample is detected (see Fig.~\ref{fig:setup}(a)(iv))~\cite{diddams_dispersion_1996}. A well-known example of a WLI in an unbalanced configuration is optical coherence tomography~\cite{huang_optical_1991}, widely used in medicine and biology. In contrast, a WLI operating in balanced configuration is commonly employed in Fourier-transform infrared spectrometry~\cite{griffiths_fourier_2007}.

\begin{figure}[ht!]
\centering\includegraphics{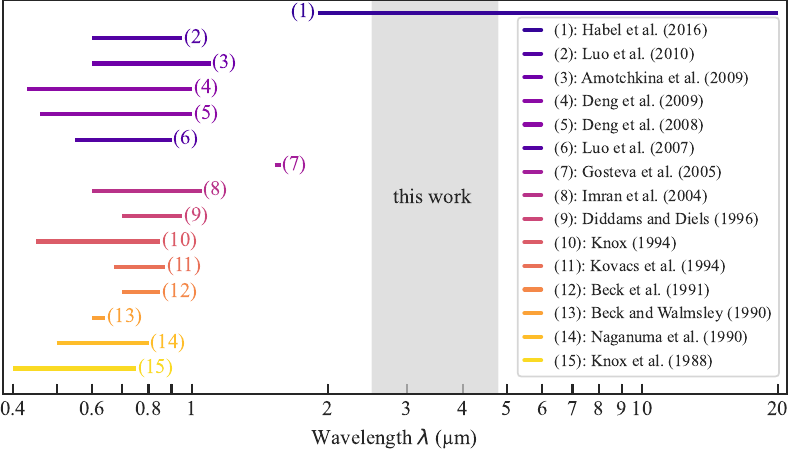}
\caption{A chronological overview of publications regarding group delay (GD), GVD, and GDD measurements. The respective spectral coverage is compared to this work (area shaded in grey).}
\label{fig:landscape}
\end{figure}

According to their mode of operation, WLIs can be divided into "spectral" and "temporal"~\cite{gosteva_noise-related_2005}. In a temporally resolved interferometer, the interferogram is recorded by a single photodiode and sampled in the time domain, leading to the well-know throughput and multiplex advantage of single-beam techniques~\cite{griffiths_fourier_2007}. This usually results in low equipment costs and allows for real-time alignment~\cite{gosteva_noise-related_2005}. On the other hand, a spectrally resolved WLI records the spectrum as a function of mirror position, relying on one or more of the following devices: monochromators, spectrometers, detector arrays or lock-in detection schemes, all of which add complexity and further costs to the setup.
Although the experimental setup is often simpler and less prone to errors in a temporally resolved WLI, a majority of publications~\cite{habel_group_2016, luo_group_2011, amotchkina_measurement_2009, deng_direct_2008, deng_wavelet-transform_2009, luo_direct_2010, kovacs_group-delay_1995, beck_group_1991,beck_measurement_1990, knox_interferometric_1988} are built as spectrally resolved WLIs, and only a small fraction~\cite{diddams_dispersion_1996, imran_measurement_2004, naganuma_group-delay_1990, gosteva_noise-related_2005} is built as temporally resolved WLIs. From the chronological overview in Fig.~\ref{fig:landscape} it is evident that most previously published experiments were performed in the visible spectral regime (VIS), with two notable exceptions: Gosteva \textit{et al.}~\cite{gosteva_noise-related_2005} investigated the near-infrared (NIR) spectral regime at telecom wavelengths (around \SI{1550}{\nano \metre}), and Habel \textit{et al.}~\cite{habel_group_2016} covered the mid-infrared (MIR). Furthermore, previously published results are lacking uncertainty estimates, with the exception of Ref.~\cite{gosteva_noise-related_2005}.

Among those studies, Habel \textit{et al.}~\cite{habel_group_2016} calls for further discussion due to its relevance to our work. They used a spectrally resolved WLI with a spectral range from \SIrange{1.9}{20}{\micro\metre} and confirmed this by measuring group velocity dispersion (GVD) of a \SI{3}{mm} bulk zinc selenide (ZnSe) window, measured in transmission. They employed a commercial Fourier-transform infrared spectrometer (Bruker Vertex 70) to record the spectrally resolved interferogram. In addition, they showed a GDD measurement of a broadband 9-layer distributed Bragg reflector (DBR), with a stopband in the MIR from \SIrange{8.5}{12}{\micro\metre}.

The present study shows the first GDD measurements of different DBRs in the MIR spectral region from \SIrange{2.5}{4.8}{\micro\metre}, among them novel monocrystalline and amorphous-crystalline hybrid supermirrors~\cite{truong_mid-infrared_2023,winkler_mid-infrared_2021}. The latter are based on substrate-transferred monocrystalline multilayer coatings fabricated using gallium arsenide (GaAs) and aluminum gallium arsenide (AlGaAs) via molecular beam epitaxy. In the case of the hybrid mirror, such monocrystalline DBRs are combined with an amorphous sub-coating, deposited via ion-beam sputtering (further details can be found in Ref.~\cite{truong_mid-infrared_2023}).
We employ a transfer-matrix method (TMM) to fit our results, leveraging different degrees of available information. For the amorphous-crystalline hybrid mirror, we utilize precise physical layers thicknesses obtained via scanning electron microscopy (values communicated via private communication with the manufacturer). In the case of the monocrystalline mirror, we assume theoretical layer thicknesses corresponding to a quarter of the designed center wavelength. For the all-amorphous mirror, no prior knowledge of the layer structure is available. Our setup is constructed as a temporal WLI, removing the necessity of an additional spectrometer. It is robust against external environmental influences, utilizes cost-efficient off-the-shelf components, and does not require an additional commercial FTIR in day-to-day operation compared to Habel \textit{et al.}~\cite{habel_group_2016}. Additionally, its capability to operate in balanced and unbalanced configuration makes it comparable to commercially available FTIRs.

In summary, we report GDD measurements of several types of DBRs in the MIR using a custom-built compact setup and describe a method for uncertainty estimation. Furthermore, we investigate a computationally efficient way for GDD calculations using a Savitzky-Golay filter and show that knowledge of the basic coating design can be used to improve measurement results considerably.

\section{Theory}\label{sec:theory}
In this study, we focus on Michelson-type WLIs utilizing a silicon carbide (SiC) Globar, i.e., typical FTIRs.
The interference pattern measured by a photodiode can be written as

\begin{equation}\label{equ:interference}
    I_{\text{int}}(\Delta \tau)   \propto \operatorname{Re}\bigg\{ \int_{- \infty}^{\infty} S(\omega) |r_{\text{ref}}| |r_{\text{sa}}| \text{exp}\big[i \phi_{\text{sa}}(\omega) - i \phi_{\text{ref}}(\omega) +  i \omega \Delta \tau \big] d \omega \bigg\},
\end{equation}
where $ S(\omega) $ is the spectral density of the light source, $|r_{\text{ref}}|$ and $|r_{\text{sa}}|$ are the reflectivity coefficients of the reference arm and sample arm respectively, and $\phi_{\text{ref}}(\omega)$, $\phi_{\text{sa}}(\omega)$ are the respective phases~\cite{gosteva_noise-related_2005}. The time delay $\Delta \tau = \tau_2 - \tau_1$ is introduced by a moving mirror. 

As we're interested in the GDD introduced by the sample, we need to extract the phase $\Delta\phi(\omega)= \phi_{\text{sa}} - \phi_{\text{ref}} $. This is done by performing a Fourier-transform on $I_{\text{int}}(\Delta \tau) $ to obtain

\begin{equation}
    \Delta\phi(\omega) = \arctan \bigg(  \frac{\operatorname{Im}\lbrace \mathcal{F}\{ I_{\text{int}}(\Delta \tau)  \} \rbrace }{\operatorname{Re} \lbrace \mathcal{F}\{ I_{\text{int}}(\Delta \tau)  \} \rbrace  }  \bigg)
\end{equation}
which can be separated into its constituents
\begin{equation}\label{equ:phase}
    \Delta\phi(\omega) = \phi_{\text{sa}} - \phi_{\text{ref}} = \phi_{\text{sample}} + \phi_{\text{sa, balanced}} - \phi_{\text{ref}},
\end{equation}
where $\phi_{\text{sample}}$ is the phase shift caused by the sample, which we are ultimately interested in. The term $\phi_{\text{sa, balanced}}$ sums all additional dispersion in the sample arm. It is evident from Equation \ref{equ:phase}, that we need to perform a calibration measurement to isolate the phase shift $\phi_{\text{sample}}$ caused by our sample.
Finally, the resulting GDD is computed as the second derivative of the corrected phase
\begin{equation}\label{equ:gdd}
    GDD = \frac{d^2 \phi_{\text{sample}}}{d \omega^2}.
\end{equation}
We note that this derivative needs to be obtained from usually noisy measurement data, a non-trivial task and active area of research~\cite{li_new_2014, chartrand_numerical_2011, van_breugel_numerical_2020}. In section~\ref{sec:Savitzky-Golay} we investigate the differentiation of Equation \ref{equ:gdd} using a Savitzky-Golay filter~\cite{savitzky_smoothing_1964}. This is a curve-smoothing technique to reduce noise by means of a moving window. Effectively, the smoothed curve is obtained by replacing each point of the original curve by a value obtained by fitting a polynomial of a given order through a symmetric interval around said point of the curve and then evaluating this polynomial, or its n\textsuperscript{th} derivative, at the points position. Hence, the main parameters controlling the amount of smoothing applied are the window length and the polynomial order used. 

When measuring the transmittance of a sample in balanced configuration (see Fig.~\ref{fig:setup}(a)(iv)), the transmitted intensity must also be corrected for the background as
\begin{equation}\label{equ:transmission}
    T = \frac{T_{\text{sample}}}{T_\text{background}}
\end{equation}
where $T_\text{background}$ represents the background measurement (see Fig.~\ref{fig:setup}(a)(iii)) and $T_{\text{sample}}$ denotes the transmitted intensity of the sample, measured in accordance to Fig.~\ref{fig:setup}(a)(iv).

\section{Experimental Methods}
\subsection{Setup}\label{sec:experimental_setup}
In our study, we use a WLI in Michelson-configuration as shown in Fig.~\ref{fig:setup}(b). The setup is separated into a reference interferometer for wavelength calibration (yellow background) and a WLI (blue background). The light source for the reference interferometer is a Helium-Neon laser (HeNe, Thorlabs HNLS008L-EC) with a center wavelength of \SI{632.8}{\nano \metre}. The WLI white light source (WLS) is a SiC Globar (Arcoptix ARCLIGHT-MIR) covering a spectral range from \SIrange{1}{25}{\um}. The light from the WLS is collimated by two off-axis parabolic mirrors (PM\textsubscript{1,2}, Thorlabs MPD129-P01) and a \SI{2}{\mm} precision pinhole (PH, Thorlabs P2000K). The beam diameter can be further adjusted with an iris in the collimated beam. For optimal modulation efficiency, we use a calcium fluoride (CaF\textsubscript{2}) 50:50 beamsplitter (BS, Thorlabs BSW511) with a wedged and AR-coated back surface. To compensate phase shifts due to the BS, an uncoated wedged CaF\textsubscript{2} compensator plate (CP, Thorlabs WG51050) is placed in the beam path of the reflected arm. The end mirrors of the reference interferometer and the WLI are placed on a voice coil based linear actuator stage (Zaber Technologies X-DMQ12P-DE52) with a travel range of \SI{12}{\mm}. This results in a maximum optical path difference of \SI{24}{\mm} corresponding to a theoretical spectral resolution of approx. \SI{0.42}{\per\cm}. The stage speed was set to \SI{15}{\mm \per \s}. The sample is placed in the fixed arm of the WLI at position M\textsubscript{S}. Since the whole setup is constructed from a \SI{30}{\mm} cage system, it is very robust and compact. Additionally, it is inherently pre-aligned, which allows us to quickly exchange the samples. The reference interferograms and WLI interferograms are captured using a Si photodiode (D\textsubscript{1}, Thorlabs PDAPC1) and a PbSe detector (D\textsubscript{2}, Thorlabs PDA 20H), respectively. The signal from the reference interferometer and the WLI are sampled using a data acquisition card (DAC, National Instruments PCI-5922) with 24 bit resolution and a maximal sample rate of \SI{15}{MS \per s}. Each of the 250 recorded double-sided interferograms consist of about 210 000 points. To exclude undesired absorption effects from atmospheric molecules, the measurements were conducted at \SI{1.3}{\millibar} using a home-built vacuum chamber.

A commercial FTIR (Bruker Vertex 80v), employing a DLaTGS detector, was used to record double-sided interferograms with a scan speed of \SI{5}{\kilo Hz} using a potassium bromide (KBr) BS and a \SI{5}{\mm} aperture size. We used the built-in software to apply a Mertz phase correction and a Norton-Beer medium apodization. The measurements were performed at \SI{2.2}{\millibar}.

\subsection{Unbalanced Configuration}\label{sec:unbalanced}
The measurement in unbalanced configuration corresponds to Fig.~\ref{fig:setup}(a)(ii), where only one arm of the interferometer interacts with the sample. As seen by Equation \ref{equ:phase}, we need to calibrate our setup. For this purpose, we made background measurements (see Fig.~\ref{fig:setup}(a)(i)), where $\text{M}_\text{S}$ and $\text{M}_\text{R}$ (see Fig.~\ref{fig:setup}(b)) are standard silver mirrors. After that, we replaced the mirror at $\text{M}_\text{S}$ with our sample and left the rest of the setup unchanged. For both, the background and the sample measurement, we recorded the interferogram and performed a Fast Fourier-transform (FFT) to get the resulting phase. The phase of the sample was corrected for the background phase in accordance with Equation \ref{equ:phase}.

\subsection{Balanced Configuration}\label{sec:balanced}
The measurement in balanced configuration (see Fig.~\ref{fig:setup}(a)(iv)), where the sample is positioned between the BS and D\textsubscript{2}, is effectively functioning as an FTIR. We used our WLI as well as the Bruker Vertex 80v to measure and compare the transmitted intensity of the sample. The measurement process is the same for both devices: first a background is recorded (see Fig.~\ref{fig:setup}(a)(iii)) and immediately afterwards the transmission of our sample. Afterwards, we applied a FFT and corrected the transmission spectrum for the background, as seen in Equation \ref{equ:transmission}.

\subsection{Simulation Model}\label{sec:simulation}
We simulated the transmission and the reflected phase of our samples by means of a TMM. An implementation in Python served as foundation of our algorithm~\cite{byrnes_multilayer_2020}. The design layer thickness of a DBR is one quarter of the designed center wavelength $\lambda_{\text{CWL}}$, which we used to simulate the monocrystalline mirror. For the amorphous-crystalline hybrid mirror we relied on layer thicknesses measured for a prior study~\cite{truong_mid-infrared_2023}. The refractive indices of GaAs and AlGaAs were taken from ~\cite{perner_simultaneous_2023}, the refractive indices of aluminium sesquioxide (Al\textsubscript{2}O\textsubscript{3}) and silica (SiO\textsubscript{2}) were taken from~\cite{kischkat_mid-infrared_2012}. For amorphous silica (a-Si) we used unpublished values provided through private communication with Dr. Valentin Wittwer from  the University of Neuchâtel.

\subsection{Data Processing}\label{sec:data_processing}
The approach for our data processing algorithm is the same for the amorphous-crystalline hybrid mirror and the monocrystalline mirror, differing only in the layer structure as explained by Truong \textit{et al.}~\cite{truong_mid-infrared_2023}. 
For the sake of brevity, we only describe the procedure for the amorphous-crystalline hybrid mirror.

In case of the unbalanced configuration (see Fig.~\ref{fig:setup}(a)(ii), we performed a nonlinear least squares fit of the TMM to our measured phase, with two free parameters, effectively scaling the respective overall thickness of the IBS-sputtered and the MBE-grown layers. The reflected phase $\phi$ is described by the function 
\begin{equation}
    \phi = \text{TMM}(\phi_{\text{measured}}, d_{\text{MBE}}^{\text{r}}, d_{\text{IBS}}^{\text{r}}),
\end{equation}
where, $d_{\text{MBE}}^{\text{r}}$ and $d_{\text{IBS}}^{\text{r}}$ represent the aforementioned scaling values of the respective layer type. The superscript r denotes reflection.

In case of the balanced configuration (see Fig.~\ref{fig:setup}(a)(iv)), the transmitted intensity $T$ is described as the function
\begin{equation}
    T = \text{TMM}(T_{\text{measured}}, d_{\text{MBE}}^{\text{t}}, d_{\text{IBS}}^{\text{t}}),
\end{equation}
where $d_{\text{MBE}}^{\text{t}}$ and $d_{\text{IBS}}^{\text{t}}$ again represent scaling values for the respective layer type. The superscript t denotes transmission.

In both cases, we obtained scaling values for the layer thickness of the MBE layers ($d_{\text{MBE}}^{\text{r}} = 1.001$ and $d_{\text{MBE}}^{\text{t}} = 1.002$) and IBS layers ($d_{\text{IBS}}^{\text{r}} =1.000$ and $d_{\text{IBS}}^{\text{t}} = 1.003$), well within the expected variations among samples from the same fabrication run~\cite{koch_thickness_2019}. These scaling values served as modifiers for the original layer thicknesses. We use a TMM based on the scaled layer thicknesses to extract the phase $\phi_\text{sample}$. The resulting GDD is computed by taking the second derivative.

\subsection{Uncertainty Estimation}\label{sec:uncertainty}
To estimate the uncertainty in the GDD, we performed a Monte Carlo-type error propagation. We measured a set of 250 interferograms and verified that the individual points, at a given optical path difference, follow a normal distribution. Then, we computed the point-wise mean and standard deviation. For each point of the averaged interferogram we selected a specific configuration of that measured point at random from a normal distribution based on the aforementioned mean value and standard deviation, generating a new randomly sampled interferogram. Next, we utilized the available measured layer thicknesses to determine the mean thickness and standard deviation of each material used. For the amorphous-crystalline hybrid mirror, the values used were as follows: GaAs with a thickness of \SI{340(1.3)}{\nano \metre}, AlGaAs at \SI{386.6(2.5)}{\nano \metre}, SiO\textsubscript{2} at \SI{822.0(6.1)}{\nano \metre}, amorphous silicon (a-Si) at \SI{294.9(3.7)}{\nano \meter}, and Al\textsubscript{2}O\textsubscript{3} at \SI{358.3(4.3)}{\nano\meter}. Based on these results, we randomly sampled the layer thicknesses and assembled a mirror stack according to the structure given by Truong \textit{et al.}~\cite{truong_mid-infrared_2023}. After creating a random interferogram and a random mirror stack, we fitted the $\phi$-TMM function - seeded with the random mirror stack and the randomly generated interferogram - as discussed in section~\ref{sec:data_processing}. We repeated this process for a total of 250 runs. The obtained errorband corresponds to the threefold standard uncertainty 3s of the fitted results.

\section{Results}\label{sec:results}
\subsection{Amorphous-Crystalline Hybrid Mirror}\label{sec:results_hybrid}
\begin{figure}[ht!]
\centering\includegraphics{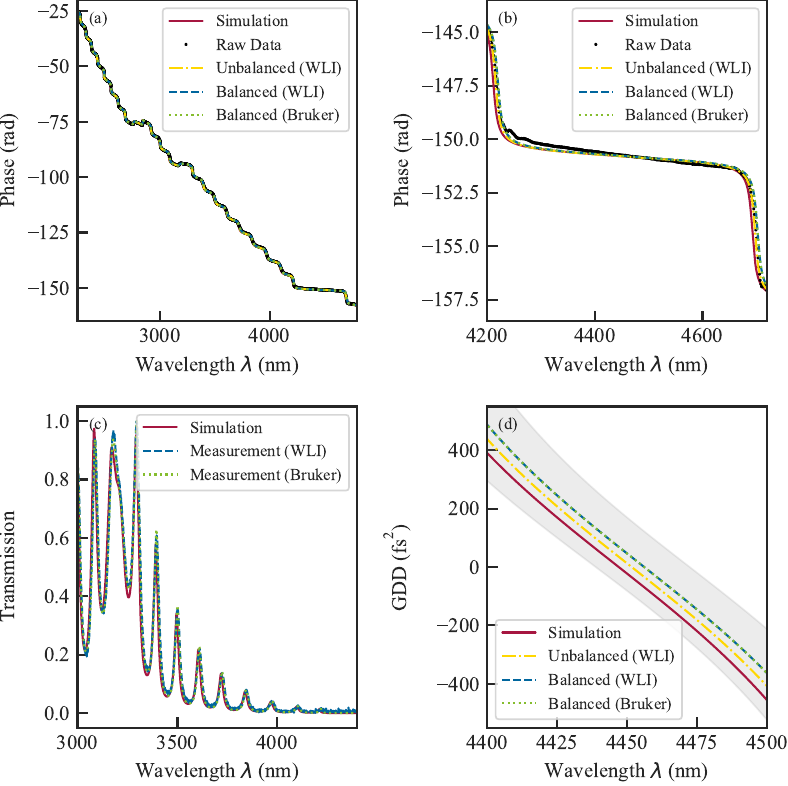}
\caption{Results for the amorphous-crystalline hybrid mirror. (a) Simulation, measured phase in unbalanced configuration (raw data), and the phases obtained from our fit procedures explained in section~\ref{sec:data_processing}, all displayed over a broad spectral range. (b) A closer look into the stopband region of (a). We observe minor irregularities at approx. \SI{4250}{\nano \metre}. (c) Comparission of the transmitted intensity in balanced configuration of our home-built WLI and the Bruker Vertex 80v, together with the simulation results using a TMM. 
(d) GDD of the amorphous-crystalline hybrid mirror. The simulation and the results of the balanced and unbalanced measurements are shown. The error band is obtained as explained in section~\ref{sec:uncertainty}, and correspond to the threefold standard uncertainty 3s.}
\label{fig:results_hybrid}
\end{figure}

The results of the amorphous-crystalline hybrid mirror is shown in Fig.~\ref{fig:results_hybrid}. In Fig.~\ref{fig:results_hybrid}(a), we show the phases obtained from the unbalanced and balanced measurements and corresponding evaluation routines (see section~\ref{sec:data_processing}). On this scale, all curves are in close agreement. A zoom into the stopband region of the amorphous-crystalline hybrid mirror (Fig.~\ref{fig:results_hybrid}(b)) reveals a deviation in the raw data in unbalanced configuration at approx. \SI{4250}{\nano \metre}. A simulation using the HITRAN~\cite{gordon_hitran2020_2022} database showed, that CO\textsubscript{2} absorption bands are in the vicinity of those deviations, suggesting the deviations result from residual absorption of CO\textsubscript{2}. However, since $\lambda_\text{CWL}$ remains unchanged and the phase irregularities occur on the stopband side, their impact on mirror quality appears to be locally. 

In Fig.~\ref{fig:results_hybrid}(c) we compare results of our custom-built WLI in balanced configuration with transmittance measurements obtained with a commercial FTIR (Bruker Vertex 80v), showing a spectral region with side lobes typical of DBR structures. We observe a shift of about \SI{4}{\nano \metre} between the two measured spectra and the simulated spectrum along the wavelength axis, which we attribute to a mismatch between the layer-thicknesses used for simulation and the real physical thickness of the layers.

The resulting GDD from all the aforementioned methods is shown in Fig.~\ref{fig:results_hybrid}(d). The GDD obtained from the two measurements in balanced configuration are in close agreement with each other. However, the GDD extracted from the unbalanced measurement is systematically lower by, on average, \SI{38}{\femto s^2} compared to the balanced measurements (see section~\ref{sec:conclusion} for details). On average, the simulation result within the stopband is \SI{40}{\femto s^2} lower than the unbalanced configuration measurement. This is evident from Fig.~\ref{fig:results_hybrid}(c), owing to the slight shift between measurement and simulation of the transmitted intensity. However, all results obtained from the various methods are within the threefold standard uncertainty of the unbalanced measurement, which is~\SI{\pm 125.7}{\square\femto \s} at $\lambda_\text{CWL}$.

\subsection{Savitzky-Golay Filter}\label{sec:Savitzky-Golay}
\begin{figure}[ht!]
\centering\includegraphics{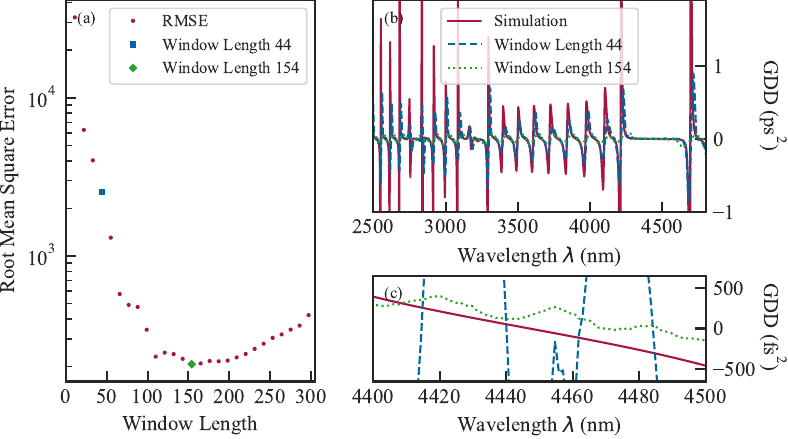}
\caption{(a) Root mean square error of the amorphous-crystalline hybrid mirror for different window lengths spanning from 11 to 301 points with a minimum at 154 points. (b) GDD along a broad wavelength range with window length equal to 44 and 154 points. (c) Zoom to the stopband of (b). At window length of 44 points we observe significant fluctuations compared to window length of 154 points. A third order polynomial was used for all calculations.}
\label{fig:savgol_hybrid}
\end{figure}

We also calculate the GDD directly, i.e., as the second derivative of the reflected phase measured in unbalanced configuration. This does not require a TMM-based fit routine, and hence no prior knowledge of the DBR design or materials. For this, we relied on a Savitzky-Golay filter~\cite{savitzky_smoothing_1964} with the main parameters being window length and polynomial order, as described in section~\ref{sec:theory}. As we are interested in the second derivative, we select a third-order polynomial. Consequently, we focus on optimizing the window length based on the results in section~\ref{sec:results_hybrid}. For that purpose, we computed the GDD of the amorphous-crystalline hybrid mirror using the Savitzky-Golay filter with increasing window length and then calculated the root mean square error (RMSE) with regards to simulated GDD within the stopband region from \SIrange{4.4}{4.5}{\micro \metre}. With a stopband width of \SI{100}{\nano \metre} and a resolution of approximately \SI{0.84}{points \per \nano \metre} within the stopband, we find an optimal window length of 154 points for reliable GDD estimation. This is illustrated in Fig.~\ref{fig:savgol_hybrid}(a), where a region of low RMSE is observed between window lengths of 100 and 200 points, with a minimum at 154 points. The derived optimal window length of 154 points translates, with the given resolution, to a spectral window length of \SI{183}{\nano \metre}, which is 1.83 times the stopband width. These values can serve as a reference for determining the optimal window length for different resolutions and stopband widths. In Fig.~\ref{fig:savgol_hybrid}(b) we see that a GDD calculated with shorter window length captures resonance peaks and is in good overall agreement with the simulated GDD and effects of oversmoothing are only visible at extreme values. However, when focusing on the stopband (Fig.~\ref{fig:savgol_hybrid}(c)), spurious fluctuations arising from a low window length become significant compared to the optimal window length. The GDD obtained from the optimal window length of 154 points is significantly smoother and serves as an excellent approximation of the simulated values.

\subsection{Monocrystalline Mirror}\label{sec:results_crystal}
\begin{figure}[ht!]
\centering\includegraphics{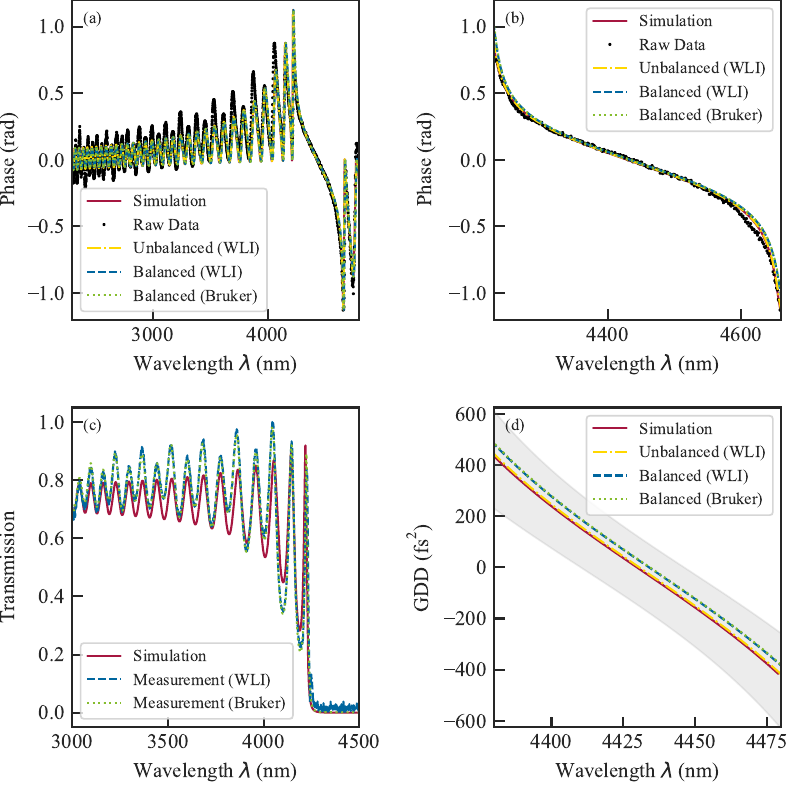}
\caption{Results for the monocrystalline mirror. (a) Simulation, measured phase in unbalanced configuration (raw data), and the phase-fit results as explained in section~\ref{sec:data_processing}. (b) A zoomed view of the stopband region of (a). We see a strong agreement of our measurement with the simulation and the fitted results. (c) The transmitted intensity measured in balanced configuration with our WLI and a Bruker Vertex 80v. (d) GDD of the monocrystalline mirror obtained from the balanced and unbalanced measurements together with the simulation result. The error band corresponds to the threefold standard uncertainty 3s, obtained as explained in section~\ref{sec:uncertainty}. }
\label{fig:results_crystal}
\end{figure}

We repeated the measurements and analysis outlined in section~\ref{sec:results_hybrid} for a monocrystalline DBR. In Fig.~\ref{fig:results_crystal}(a) we show the results for the phase over the complete measurement range for the various methods. Based on Fig.~\ref{fig:results_crystal}(a) and (c), we see a variation of transmission as well as phase values in the raw measurement, when compared to the simulation and the fitted values. This is attributed to non-uniform layer thicknesses of the sample. Within the stopband (Fig.~\ref{fig:results_crystal}(b)), the simulation, the measurement in unbalanced configuration, and our fit results are in excellent agreement. Likewise, the results for the measurements with our WLI in balanced configuration, depicted in Fig.~\ref{fig:results_crystal}(c), agree well with those obtained with the commercial FTIR. Deviations of the measurements from the simulations, similar to those observed in \ref{fig:results_crystal}(a), are well-explained by layer thickness deviations of the as-fabricated mirror from a perfect quarter-wave design. As can be seen in Fig.~\ref{fig:results_crystal}(d), the GDD resulting from the measurement in balanced configuration and the commercial FTIR are virtually identical. This is expected, as they are based on closely matching transmission spectra (see Fig.~\ref{fig:results_crystal}(c)). We note that the GDD resulting from the measurement in unbalanced configuration within the stopband is, on average, \SI{33}{\femto s^2} below the GDD resulting from the balanced configuration, and about \SI{2}{\femto s^2} higher than the simulation, similar to what we observe for the amorphous-crystalline hybrid mirror in section~\ref{sec:results_hybrid}. A discussion about this behavior is provided in section~\ref{sec:conclusion}. All results are within the threefold standard uncertainty of \SI{\pm 134.2}{\square\femto \s} at $\lambda_\text{CWL}$ obtained as described in section~\ref{sec:uncertainty}.

\subsection{All-amorphous Mirror}
\begin{figure}[ht!]
\centering\includegraphics{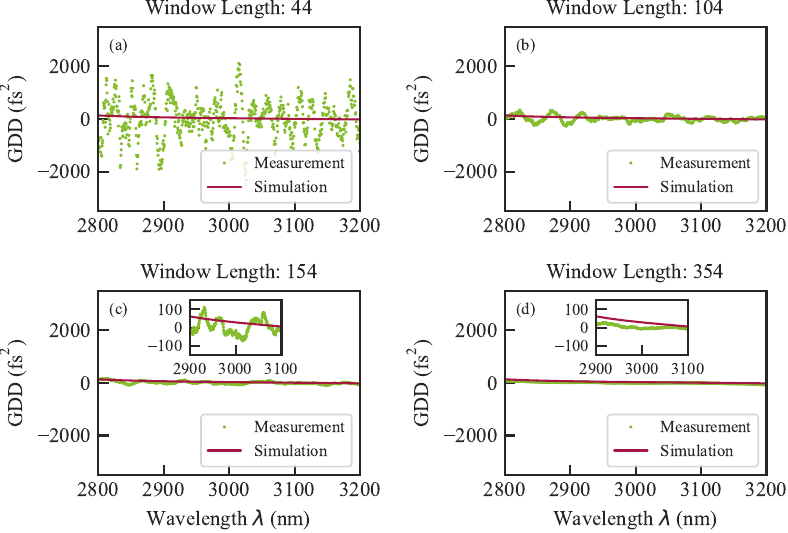}
\caption{Results for the all-amorphous mirror. (a)-(d) Comparison of GDD calculated by the Savitzky-Golay filter using different window lengths. The fluctuations reduce with bigger window length until the point where we introduce a systematic error due to large window lengths. This systematic error manifests into a shift of the GDD along the y-axis.}
\label{fig:IBSSavGolWindow}
\end{figure}

In this part of the study, we investigate the GDD of an all-amorphous commercial mirror with a center wavelength of approx. \SI{3000}{\nano \metre} without any prior knowledge of the mirror in terms of layer materials or design. However, based on the transmittance spectrum and trial-and-error, we see good agreement between the measured spectrum and a TMM using a quarter-wave design with 14 perionds based on ZnSe and Ytterbium(III) oxide. This model was used to simulate the GDD within the stopband. We also used the results from section~\ref{sec:Savitzky-Golay} to approximate the GDD measured with our WLI using the Savitzky-Golay filter and compare it to the simulation, again using different window lengths to demonstrate the impact of the Savitzky-Golay filter on the GDD. Fig.~\ref{fig:IBSSavGolWindow}(a) shows the result for a window length of 44 points, where significant fluctuations are visible. Increasing the window length gradually (Fig.~\ref{fig:IBSSavGolWindow}(b)-(d)), we observe the expected smoothening behavior and increasing precision. At a window length of 154 points, which we derived as an optimum for the amorphous-crystalline hybrid mirror, we see a significant reduction of random fluctuations without signs of oversmoothing. We note that this holds true even though the optimum window length was derived for a slightly different stopband geometry. This aligns with the findings in Fig.~\ref{fig:savgol_hybrid}(a), which illustrates that close-to-optimal results can be obtained at window lengths around 100 to 200 points. By increasing the window length further, the fluctuations decrease more (Fig.~\ref{fig:IBSSavGolWindow}(d)), but the GDD in general is shifted downwards, which we identify as first signs of oversmoothing. From this we infer that using a Savitzky-Golay filter always requires a compromise between precision and computationally introduced systematic errors based on the chosen window length.

\section{Discussion and Conclusion}\label{sec:conclusion}
In summary, we report GDD measurements for monocrystalline and amorphous-crystalline hybrid mirrors, as well as an all-amorphous Bragg reflector, covering a spectral range from \SIrange{2.5}{4.8}{\micro \metre}. We obtained these measurements with a custom-built and user-friendly WLI which can be used in balanced and unbalanced configuration and is easily adapted to cover an even broader wavelength range. As noted in section~\ref{sec:results_hybrid} and \ref{sec:results_crystal}, we observe that the GDD obtained in balanced configuration is systematically higher compared to the measurement in unbalanced configuration for both, the  amorphous-crystalline hybrid mirror and the monocrystalline mirror. We verified this by repeating the balanced measurement with a commercial FTIR (Bruker Vertex 80v). This systematic offset may be attributed to several factors, including but not limiting to, subtle variations in the angle of incidence of the white light between the two measurement configurations, a factor particularly significant given the sample's placement on a curved substrate. However, this discrepancy lies comfortably within the bounds of the uncertainty estimation we conducted, hence supporting the reliability of our results. An uncertainty analysis based on a Monte Carlo-style approach was performed, suggesting that the observed systematic deviations are well within the 3s uncertainty. Additionally, we conducted a detailed investigation of direct numerical differentiation using a Savitzky-Golay filter. We could show that there is an optimal window length dependent on the stopband width and spectral resolution of the measurement, at which the directly derived GDD in the stopband agrees well with simulated values. This window length was used to determine the GDD of an all-amorphous broadband DBR with center wavelength at \SI{3}{\micro \metre}. We demonstrated the approach's robustness by using results at \SI{4.5}{\micro \metre} to obtain a consistent GDD measurement at \SI{3}{\micro \metre}. Furthermore, we derived a formula to determine an optimal window length to obtain reliable GDD estimates within a mirror's stopband using a Savitzky-Golay filter with a third-order polynomial. For this purpose, only the knowledge of spectrometer resolution and stopband width is necessary, as can be obtained from a simple transmittance measurement or even manufacturer specifications. Those findings allow for a fast determination of the GDD by a single measurement of the reflected phase without the need of additional equipment or simulations. 

\section*{Acknowledgements} 
This research was funded in whole or in part by the Austrian Science Fund (FWF)[10.55776/P36040]. For open access purposes, the author has applied a CC BY public copyright license to any author accepted manuscript version arising from this submission. The financial support by the Austrian Federal Ministry of Labour and Economy, the National Foundation for Research, Technology and Development and the Christian Doppler Research Association is gratefully acknowledged.\\
We thank Dido Denier van der Gon (University of Vienna) for assistance with the Bruker Vertex 80v, Dr. Valentin Wittwer (University of Neuchâtel) for discussions and data on IBS materials, and Dr. Garrett Cole, Dr. Gar-Wing Truong (Thorlabs Crystalline Solutions), and Dr. Adam J. Fleisher (NIST) for providing the monocrystalline and amorphous-crystalline samples. We also thank Martin Haberl for welding the vacuum chamber. Instrument identification in this paper is solely for experimental clarity and does not imply endorsement. Artificial intelligence tools were used for grammar checking and typesetting to enhance readability.

\section*{Conflict of Interest}
The authors declare no conflicts of interest.

\section*{Data Availability Statement}
The data that support the findings of this study are available from the corresponding authors upon reasonable request.

\bibliography{references_gdd}

\end{document}